\documentclass[11pt]{article}

\usepackage[final]{acl}

\usepackage{times}
\usepackage{latexsym}

\usepackage[T1]{fontenc}

\usepackage[utf8]{inputenc}

\usepackage{microtype}

\usepackage{inconsolata}

\usepackage{graphicx}
\usepackage{booktabs}
\usepackage{multirow}
\usepackage{subcaption}
\usepackage{xcolor}
\graphicspath{{figures/}}
\usepackage{tabularx}
\usepackage{amsmath}

\newcommand{\added}[1]{#1}

\title{Probing Large Audio-Language Models for Compositional Understanding of Sounding Actions}

\author{
\textbf{Michel Olvera} \hspace{1.5em} \textbf{Paraskevas Stamatiadis} \\
\textbf{Changhong Wang} \hspace{1.5em} \textbf{Gaël Richard} \hspace{1.5em} \\
LTCI, Télécom Paris, Institut Polytechnique de Paris\\
\texttt{\{olvera,stamatiadis\}@telecom-paris.fr}
}

\usepackage{fancyhdr}
\usepackage{ragged2e}  
\fancypagestyle{firstpage}{
    \fancyhf{} 
    \setlength{\footskip}{28pt}  
    \fancyfoot[C]{\centering \footnotesize{\copyright 2026 Association for Computational Linguistics. Licensed under CC BY 4.0.}} 
}

\begin{document}
\maketitle

\thispagestyle{firstpage}

\begin{abstract}
Large audio-language models (LALMs) excel at understanding and reasoning tasks over atomic sound events, yet their ability to infer higher-level human activities from such fine-grained events remains largely unexamined. Everyday human actions and activities, such as setting a table, cleaning the house, or preparing a breakfast emerge compositionally from temporally distributed sound events, requiring abstraction beyond the event-centric granularity that dominates current training and evaluation paradigms. Our benchmark evaluates a wide set of LALMs under a principled framework that tests how language-based reasoning, grounded in acoustic perception, structures sound abstractions into higher-level understanding. By systematically varying exemplar typicality and distractor similarity, our evaluation exposes \added{that current models do not reliably perform compositional inference from atomic acoustic events to higher-level human activities solely from audio.} All data, taxonomies, and evaluation scripts are publicly available on our companion website: \small{\url{https://alm-sounding-actions.onrender.com/}}.
\end{abstract}

\section{Introduction}

\begin{figure}[t]
  \centering
  \includegraphics[width=\columnwidth]{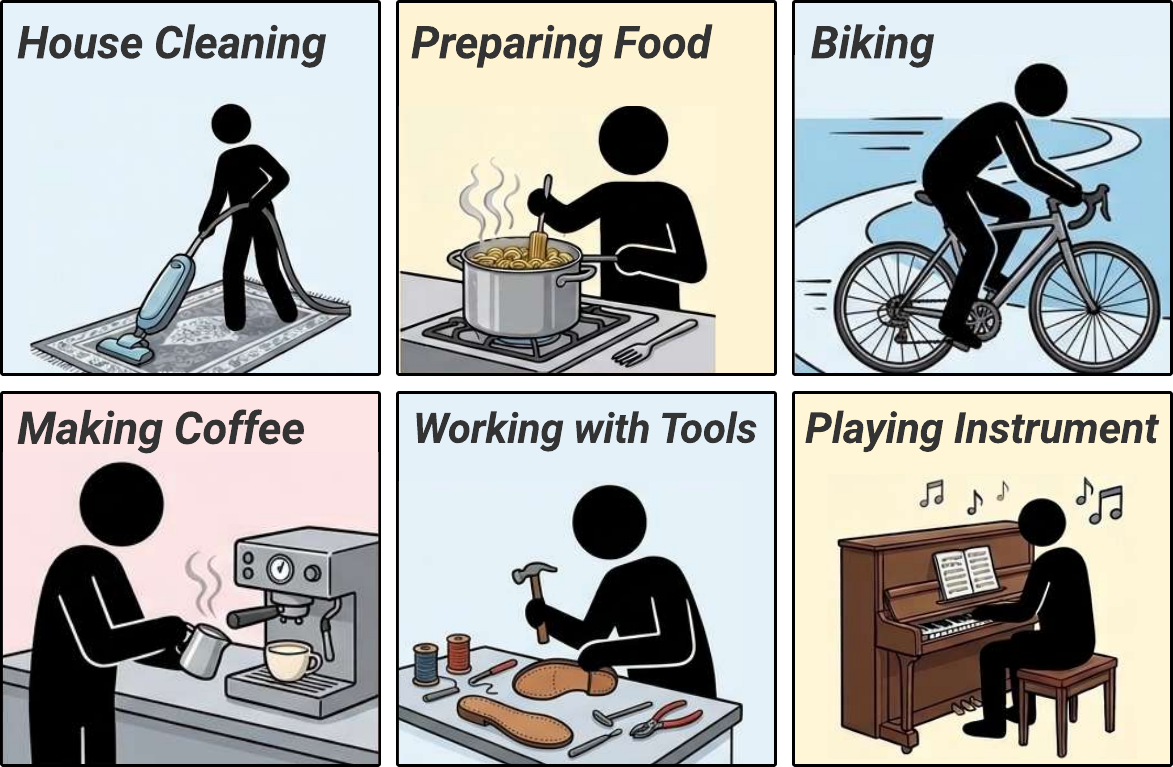}
  \caption{%
    \textbf{Sounding activities.}
    Examples of everyday human activities inferred from compositional sequences of acoustic events.
  }
  \label{fig:sounding_activities}
\end{figure}

Human auditory perception naturally operates at multiple levels of abstraction. Beyond recognizing isolated sounds such as a door slam, a keyboard click, or a dog bark, humans continuously integrate streams of acoustic events into coherent interpretations of everyday activities and behaviors. The sound of running water followed by cupboard movements and metallic clattering may immediately suggest someone preparing a meal; the alternating rhythm of footsteps, keys jingling, and a door opening may signal arriving home. These activities are not defined by a single acoustic event, but rather emerge compositionally from sequences of temporally distributed and loosely coupled sound cues (Figure \ref{fig:sounding_activities}). While such abstraction is effortless for humans, enabling machines to infer high-level actions from sound remains a fundamental challenge. 

Over the past years, multimodal models grounding audio and language have achieved remarkable progress in recognizing atomic acoustic events and short-duration sound scenes. Large-scale contrastive pretraining has enabled models to associate audio with rich textual descriptions, yielding strong performance on sound classification, retrieval, and open-vocabulary audio understanding tasks. However, existing evaluation benchmarks remain predominantly centered on isolated events and acoustically explicit categories, such as \textit{dog barking}, \textit{glass breaking}, or \textit{vacuum cleaner}. Even when multiple sound sources co-occur, the underlying task often reduces to identifying constituent events rather than inferring the higher-level human activity they collectively imply. As a result, current benchmarks provide limited insight into whether audio-language models can reason compositionally over sound and map collections of atomic events into semantically meaningful human actions. This leaves a fundamental open question for real-world auditory understanding:

\textit{Can large audio-language models identify human activities from compositions of acoustic events by reasoning over what they hear?}

To answer this question, we take a systematic step in this direction by introducing a benchmark for compositional understanding of sounding human actions and activities. Curated from large-scale egocentric and exocentric human activity datasets, the benchmark organizes approximately one hundred audio-identifiable actions and activities into a hierarchical taxonomy spanning domestic routines, tool usage, sports, mobility, entertainment, and social interaction. To evaluate compositional reasoning systematically, we design a controlled probing framework along two complementary dimensions of difficulty — exemplar typicality and distractor similarity — and across four complementary protocols: closed-set recognition, open-ended prediction, event-grounded chain-of-thought, and hierarchical classification.

Our experiments reveal a substantial gap between atomic sound recognition and compositional auditory understanding. We hypothesize that current LALMs, despite their elevated reasoning capabilities, remain anchored to the granularity of their training signal \added{and do not reliably exhibit compositional abstraction over acoustic events in their outputs, regardless of the underlying mechanism}. Models frequently confuse acoustically overlapping routines such as cooking, cleaning, and organizing, and struggle particularly on atypical exemplars that require robust abstraction beyond prototypical sound patterns. These findings suggest that compositional auditory understanding constitutes a distinct capability not yet captured by existing training paradigms or benchmarks.

In summary, we present the following contributions:
(1) The first benchmark \added{targeting} compositional understanding of sounding human actions and activities \added{from egocentric acoustic event sequences}, organized around a hierarchical taxonomy of audio-identifiable everyday behaviors.
(2) A principled probing framework spanning closed-set, open-ended, event-grounded, and hierarchical evaluation protocols.
(3) A comprehensive evaluation of state-of-the-art LALMs, revealing fundamental limitations in generalizing from atomic acoustic events to semantically structured human activities.
To facilitate future research, all datasets, prompts, model outputs, and codebase resources are hosted on our companion website.\footnote{\url{https://alm-sounding-actions.onrender.com/}}

\section{Related Work}

\paragraph{Audio-Language Models (ALMs)}
Audio-language modeling has progressed from contrastive \added{audio-text representation learning} \citep{guzhov2022audioclip, elizalde2023clap, wu2023large} to Large Audio-Language Models (LALMs) that couple audio encoders with LLMs for open-ended audio understanding and reasoning \citep{deshmukh2023pengi,gong2023listen,chu2023qwen,tang2023salmonn,kong2024audio, ghosh2025audio, goel2025audio,Qwen2.5-Omni,li2025baichuan}. More recently, several works have explicitly targeted complex reasoning through chain-of-thought strategies, structured inference, and semantic decomposition \citep{xie2025audio,ma2025audio,li2025reinforcement,wijngaard2025audsemthinker} yielding gains on tasks requiring multi-step inference over acoustic events. However, whether structured reasoning generalizes to compositional auditory understanding i.e., inferring structured human activities from sequences of acoustic events, remains unexamined. Our work directly targets this gap by probing whether models can abstract acoustic event sequences into the semantically coherent human activities they jointly imply. 

\paragraph{Benchmarks for Audio Understanding and Reasoning}
\added{Recent benchmarks, including CompA \citep{ghosh2023compa}, MMAU \citep{sakshi2025mmau}, MMAU-Pro \citep{kumar2026mmau}, and MMAR \citep{ma2025mmar}, evaluate various aspects of audio understanding, compositional reasoning, and multimodal capabilities. However, these benchmarks contain scarce activity-related examples as their primary focus is broader auditory perception and general-purpose reasoning rather than a systematic evaluation of inferring higher-level human activities from compositions of atomic acoustic events. In particular, none of these benchmarks isolates activity typicality or distractor similarity as controlled difficulty factors. Our benchmark complements these efforts by isolating this specific capability through controlled variation along these two axes.}

\paragraph{Sound Event and Activity Recognition}
Sound event detection and audio tagging have been shaped by benchmarks such as AudioSet \cite{elisAudioset}, FSD50K \cite{fonseca2022FSD50K}, and ESC-50 \cite{piczak2015dataset}, which define the dominant evaluation paradigm: short, isolated clips annotated with atomic event labels. Models trained on these sources, including large-scale audio tagging systems, excel at recognizing individual sound categories but are not designed to infer structured human activities from co-occurrences of such event sequences. Audio-visual datasets such as VGGSound \cite{Chen20} extend event recognition to a larger scale but retain the clip-level, single-label framing. Large-scale egocentric corpora such as Ego4D~\cite{Grauman_2022_CVPR}, Ego-Exo4D~\cite{Grauman_2024_CVPRexo} and EPIC-KITCHENS \cite{Damen2020Collection} offer temporally extended, activity-grounded recordings of daily human behavior and have driven major advances in visual activity recognition. Despite the richness of their audio streams, these datasets have not been used to probe whether models can perform compositional auditory inference: recognizing what a person is doing by making sense of acoustic cues rather than identifying isolated events. Closest to our setting, \citet{demirel2025using} use LLMs to fuse audio and motion sensor streams for zero-shot activity classification on a subset of Ego4D, but as sensor fusion rather than probing compositional reasoning over acoustic events alone. Our benchmark is the first to formalize this capability as a structured evaluation target for LALMs.

\section{Methodology}
The methodology proceeds in three stages: (i) constructing a hierarchical taxonomy of sounding actions, (ii) building a controlled dataset of audio exemplars, and (iii) defining the factorial structure of the benchmark.
\subsection{Taxonomy of Sounding Actions}
\label{subsec:taxonomy}
To probe compositional reasoning in ALMs, we require a structured label space of human activities as a principled basis for every downstream experimental design decision.

\paragraph{Source datasets.}
We consolidate activity annotations from large-scale egocentric and exocentric datasets covering everyday human behaviour: Ego4D, Ego-Exo4D and HD-EPIC~\cite{perrett2025hdepic}.
These corpora are selected for three specific reasons. First, they provide annotated recordings describing what a person is doing, rather than event-centric tags describing what is heard in isolation. This matches the compositional target of our benchmark. Second, their recording conditions provide acoustic variability that reflects real-world listening conditions. \added{Third, to our knowledge, these datasets are not part of the documented pretraining corpora of the evaluated open-source models. For proprietary models, pretraining data are undisclosed and cannot be independently verified.} We rely on SALT~\citep{Stamatiadis2024} to standardize annotations across these datasets.

\paragraph{Audio-identifiability filtering.}
Source datasets contain activities that are not all distinguishable from audio alone. We seek to retain activity labels whose constituent sound events support reliable recognition (\textit{e.g., house cleaning, playing musical instrument, metal working}). We exclude both acoustically silent activities, whose incidental sounds are non-diagnostic (\textit{e.g., reading}), and activities whose recognition relies primarily on visual cues (\textit{e.g., cooking pasta}).
We implement this criterion using a two-round selection protocol. First, we assess whether an activity is, in principle, recognizable from sound. Second, we verify distinctive and audible acoustic cues in sampled clips. Authors cross-validated these assessments, and we include only labels that passed both rounds with full consensus. Full rubrics and decision criteria are provided in Section \ref{sec:audibility_filtering}.

\paragraph{Hierarchical structure.} The resulting taxonomy comprises 99 nodes organized across a maximum of five abstraction levels. The hierarchy is derived from the activity label spaces of the source datasets, which we taxonomically organize by grouping labels under progressively coarser semantic categories. Each node is placed within a strict parent–child hierarchy: for example, \textit{floor sweeping} $\rightarrow$ \textit{household cleaning} $\rightarrow$ \textit{domestic activities}.

\subsection{Audio Exemplar Curation}
\label{subsec:dataset}

\paragraph{Segment extraction.}
Source data is accompanied by time-stamped activity labels. We extract segments solely from test recordings where a single label is active, ensuring that no segment mixes multiple activity classes, thereby framing the benchmark as a single-label classification task. We keep segments between 5–10 seconds long, deemed sufficient to capture the multiple sounds that define an activity. Each segment retains its original activity label. This process left us with effective data for 85 of the 99 taxonomy labels; 14 labels had too little data after segment extraction \footnote{No clips from HD-EPIC dataset were selected after such procedure, its contribution is thus limited to the label space of the taxonomy.}.

\paragraph{Embedding prototype computation.}
We encode segments into audio embeddings using 
CLAP~\cite{elizalde2023clap}, a model trained on millions of audio-language pairs whose geometry reflects semantic, human-interpretable distinctions
between sound categories.
For a given label $c$, we compute a class prototype $\mu_c$ as the average of all segment embeddings assigned to that label.
This prototype represents an acoustic summary of what the activity \textit{``sounds like''}.

\paragraph{Exemplar selection.}
For each activity class, we select exemplars based on their $\ell_2$ distance to the class prototype $\mu_c$ in the CLAP embedding space.
Two sets are constructed per label: \emph{\textbf{(a)~central exemplars}}: segments whose embeddings lie closest to $\mu_c$, representing canonical, prototypical instances of the activity; \emph{\textbf{(b)~peripheral exemplars}}: segments whose embeddings lie farthest from $\mu_c$ within the class, after trimming extreme outliers (segments beyond two standard deviations above the mean intra-class distance to $\mu_c$), representing acoustically atypical instances of the same activity.

To ensure diversity we sample one segment per recording, ensuring no single recording dominates either set. Under this constraint, we cap selection at 20 central and 20 peripheral exemplars per label, though class imbalance in the source datasets means some labels fall short of this target. Any label with less than five central clips after this process is excluded as a ground-truth candidate, but may still appear as a distractor label, whose construction is explained below. 

\paragraph{Distractor selection.}
For a given ground-truth label, three distractor labels are drawn either from \emph{\textbf{(a)~distant classes}}: the 10 classes whose prototypes are farthest from the ground-truth prototype or \emph{\textbf{(b)~near classes}}: the 10 classes whose prototypes are closest. Labels in a direct ancestor-descendant relationship with 
the ground-truth are excluded to prevent trivial taxonomic confounds. Distractors are sampled independently for each exemplar, used to compose a four-option forced-choice question per item.

\paragraph{Dataset statistics.}
\added{
The final benchmark comprises 85 labels and 1,938 clips (903 central, 1,035 peripheral), with a median of 35 clips per label (range 6–40). Most labels are near-balanced between central and peripheral exemplars; however, 19 labels have fewer than 20 total clips, 4 have fewer than 10, and 8 show strong central/peripheral skew ($\geq$5:1 or one-sided, e.g., \emph{drinking} 0/6, \emph{counting money} 7/0). We flag these low-support labels as higher-uncertainty cases in the released metadata.}

\subsection{Benchmark Factorial Structure}
\label{subsec:difficulty_axes}
\begin{figure}[t]
  \centering
  \includegraphics[width=\columnwidth]{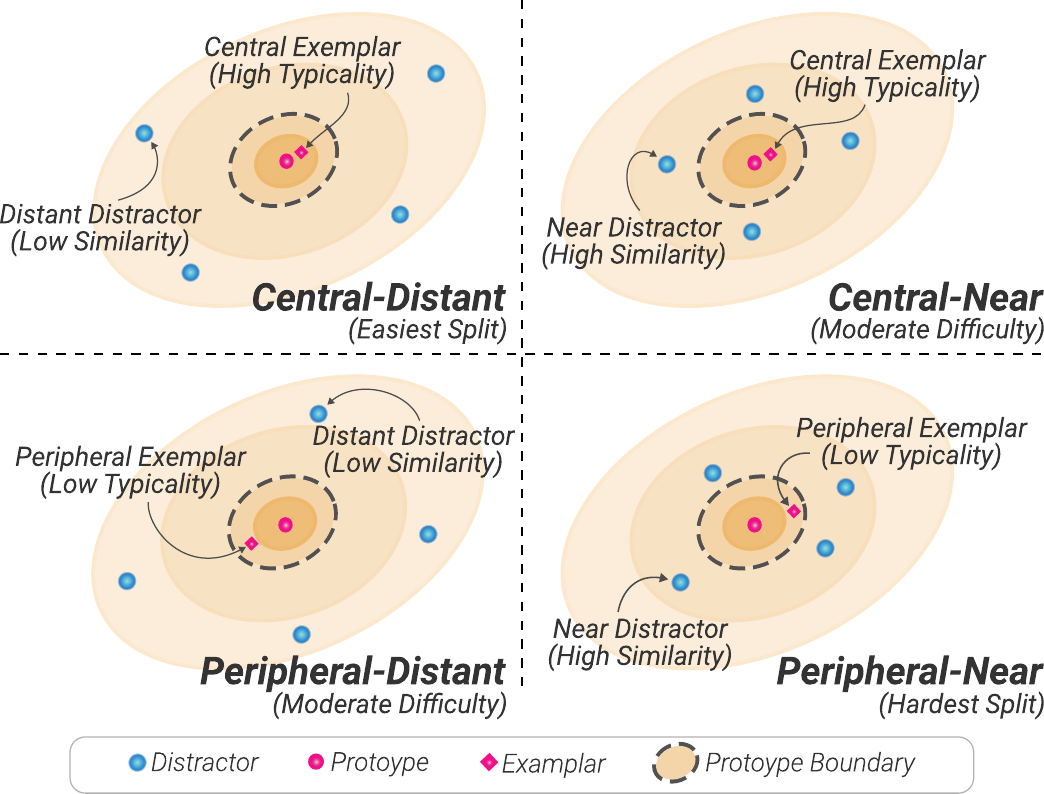}
  \caption{%
    Combining exemplar typicality (central vs. peripheral) and distractor similarity (distant vs. near) yields four conditions of varied difficulty, from Central–Distant (easiest) to Peripheral–Near (hardest).
  }
  \label{fig:cross_factor_splits}
\end{figure}

To control task difficulty, we structure our evaluation around two independent probing factors:

\paragraph{Factor 1 — Exemplar typicality.}
Reflects robustness to within-class variation by contrasting central exemplars (high typicality) with peripheral exemplars (low typicality). This factor controls \emph{audio representativeness} and tests whether a model's internal representation of an activity is sufficiently robust to recognize either canonical or non-canonical instantiations.

\paragraph{Factor 2 — Distractor similarity.}
Reflects precision in inter-class discrimination by contrasting distant distractors (low similarity to the target) with near distractors 
(high similarity to the target). This factor tests whether a model can resolve between acoustically similar or dissimilar activities. 

Crossing these factors (Figure \ref{fig:cross_factor_splits}) yields four conditions each reflecting a different level of difficulty: \emph{\textbf{(1)~Central–Distant}}: canonical exemplars, acoustically dissimilar distractors. This condition is expected to be the easiest; \emph{\textbf{(2)~Central–Near}}: canonical exemplars, acoustically similar distractors. It probes fine-grained inter-class discrimination; \emph{\textbf{(3)~Peripheral–Distant}}: less typical exemplars, dissimilar distractors. It probes robustness to non-canonical inputs in the absence of confounding alternatives; \emph{\textbf{(4)~Peripheral–Near}}: less typical exemplars, similar distractors. The hardest split, requiring simultaneous within-class invariance and fine-grained inter-class discrimination.

 
\section{Experimental Setup}
\label{sec:experimental_setup}
We evaluate audio-language models' compositional inference capabilities through structured probe families, across a diversified prompt bank (see Appendix~\ref{app:prompt_bank}), and comprehensive evaluation metrics. To ensure reproducibility, our prompt bank, raw model responses, and evaluation framework are distributed on our companion website. 
\subsection{Models}
\label{subsec:models}
We evaluated approximately 30 audio-language models (ALMs) which we grouped into four categories: \textit{open-source generative}, \textit{instruction-tuned}, \textit{reasoning-augmented}, and \textit{proprietary frontier models}. The full model list is show in Appendix \ref{app:evaluated_models}. We also include, where available, both thinking (chain-of-thought at inference time) and no-thinking variants from the same model family to isolate the effect of test-time reasoning.
 
\subsection{Probe Family I — Flat Compositional Inference}
\label{subsec:direct_eval}
This probe family tests whether models can infer human activities from acoustic evidence without hierarchical semantic constraints. Protocols P1 and P2 evaluate compositional inference under a closed-set formulation, where models must distinguish the target activity from a controlled contrast set. Protocol P3 extends the evaluation to an open-ended setting without predefined candidate labels.
 
\paragraph{P1 — Direct Activity Inference. }
P1 tests implicit end-to-end inference in a forced-choice setting. Given an audio clip and a four-label contrast set (ground-truth activity plus three distractors), the model must return exactly one label. This protocol assesses whether compositional auditory understanding emerges without explicit reasoning scaffolds.
 
\paragraph{P2 — Event-Grounded Inference.}
P2 tests whether models can ground activity inference in explicit event recognition within a single forward pass. The model must: (i)~identify the salient sound events in the clip, and (ii)~ infer the most likely activity based on such events. This protocol probes whether eliciting event recognition improves compositional inference.
 
\paragraph{P3 — Open-Ended Activity Inference.} The model must infer the activity as a free-form noun phrase (up to five words) without any contrast set. This protocol probes unconstrained semantic grounding, lexical generalization, and alignment with the taxonomy.
 
\subsection{Probe Family II — Hierarchical Abstraction}
\label{subsec:hierarchical_eval}
This probe family tests whether models can reason consistently across semantic abstraction levels via parent-child relations. For each clip, both a coarse semantic category (\textit{e.g., domestic activities}) and a fine-grained activity node (\textit{e.g., floor sweeping}) are defined according to the taxonomy. We evaluate two protocols:
 
\paragraph{H1 — Joint Hierarchical Abstraction.}
H1 tests hierarchical abstraction in a single forward pass. Given an audio clip and candidate pools for both hierarchy levels, the model must jointly predict: (i)~a coarse semantic category, and (ii)~a fine-grained activity label.
The model must output both predictions in a fixed format (e.g., \texttt{Coarse: <label>; Fine: <label>}). This protocol probes whether models can maintain inter-level consistency while performing fine-grained compositional inference.
 
\paragraph{H2 — Sequential Hierarchical Abstraction.}
H2 explicitly decomposes hierarchical abstraction into two sequential calls: (C1)~predict the appropriate coarse semantic category, and (C2)~select the fine-grained activity label conditioned on the coarse category from C1. This protocol probes whether separating coarse and fine-grained reasoning improves hierarchical compositional inference. 
 

\subsection{Evaluation Metrics}
\label{subsec:metrics}
\begin{table}[t]
\centering
\small
\setlength{\tabcolsep}{5pt}
\begin{tabular}{lccc}
\toprule
\textbf{Protocol} & \textbf{Items/split} & \textbf{Prompts} & \textbf{Evaluations/split} \\
\midrule
P1 & 420 & 15 & 1,680 \\
P2 & 140 & 5  & 560 \\
P3 & 270 & 10 & 540 \\
H1 & 140 & 5  & 560 \\
H2 & 130 & 5  & 520 \\
\bottomrule
\end{tabular}
\caption{Eval set sizes per protocol/cross-factor split.}
\label{tab:evaluation-set-sizes}
\end{table}

We report accuracy as the primary evaluation metric across all protocols. Table \ref{tab:evaluation-set-sizes} summarizes the number of items and total evaluations for each protocol and cross-factor split. Correctness is determined using a constrained LLM-based evaluation framework. Specifically, Meta-Llama-3.1-8B-Instruct~\citep{grattafiori2024llama} assigns a binary \textit{Correct}/\textit{Incorrect} judgement to each prediction conditioned on its ground-truth label.

For closed-set protocols (P1, P2, H1, H2), predictions are evaluated conservatively against the predefined activity vocabulary, while tolerating minor formatting or typographical deviations. For hierarchical protocols, we additionally report \textit{coarse}, \textit{fine}, and \textit{joint} accuracies corresponding to correctness at different taxonomy levels.

For open-ended protocols (P3), evaluation is based on semantic equivalence rather than exact lexical identity, allowing valid paraphrases and semantic reformulations of the target activity.

To validate the evaluation procedure, we compare LLM-based judging against a deterministic lexical baseline using normalized fuzzy string matching (RapidFuzz threshold = 0.8). Both methods produce near-identical results across closed-set protocols, indicating that the LLM judge primarily reduces penalties arising from formatting mismatches in otherwise semantically correct predictions, without materially inflating performance.

\added{Judges are validated against four human annotators on 200 sampled pairs: for P3, judge-annotator agreement ($\kappa=0.745$) exceeds the human-human baseline ($\kappa=0.714$), with 96.8\% agreement ($\kappa=0.90$) against majority human labels; for closed-set protocols, agreement is 90.2\% ($\kappa=0.803$).}

\section{Results}
\label{sec:results}

\subsection{Flat Compositional Inference}
\label{subsec:results_scenario1}

\begin{figure}[t]
  \centering
  \includegraphics[width=\columnwidth]{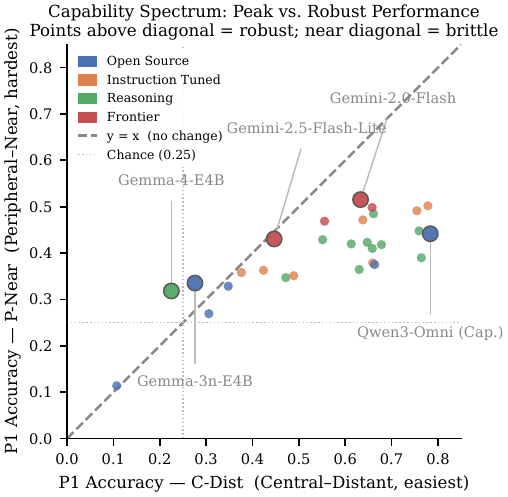}
    \caption{\textbf{Typicality robustness across model capability.} Each point plots accuracy on the easiest condition (C-Dist, central exemplars) against the hardest (P-Near, peripheral exemplars). Most models fall below the diagonal, revealing pervasive typicality sensitivity largely independent of overall capability.}
  \label{fig:p1_typicality}
\end{figure}

\paragraph{Effect of exemplar typicality.}
To isolate the effect of typicality, we compare Central-Distant against Peripheral-Distant while holding distractor similarity fixed. As shown in Figure~\ref{fig:p1_typicality}, this shift consistently induces a marked performance drop across models, confirming that atypical sound instances are intrinsically harder even under minimal acoustic confusability. This establishes typicality as a primary factor shaping task difficulty, independent of distractor structure.

\paragraph{Typicality robustness decoupled from model capability. }
As shown in Figure~\ref{fig:p1_typicality}, most models lie below the diagonal, indicating pervasive sensitivity to typicality shifts that is largely independent of overall capability. Qwen3-Omni Captioner exhibits the strongest degradation ($\Delta \approx -0.34$), achieving the highest C-Dist accuracy but only moderate generalization to P-Near, suggesting reliance on prototypical matching that breaks under atypical instances. Gemini-2.0-Flash shows a smaller but consistent drop ($\Delta \approx -0.12$), while Gemini-2.5-Flash-Lite is nearly invariant to typicality ($\Delta \approx -0.02$), emerging as the most stable model in terms of robustness within the frontier cluster despite intermediate absolute accuracy.

Interestingly, Gemma-4-E4B and Gemma-3n-E4B invert this trend ($\Delta > 0$), performing better on peripheral than central exemplars; however, this occurs in a low-accuracy regime given their speech-centric design. Overall, robustness to typicality is not monotonic with capability and can even reverse within model families, suggesting that it should be treated as an independent evaluation dimension rather than a byproduct of overall performance.

\begin{figure}[t]
  \centering
  \includegraphics[width=\columnwidth]{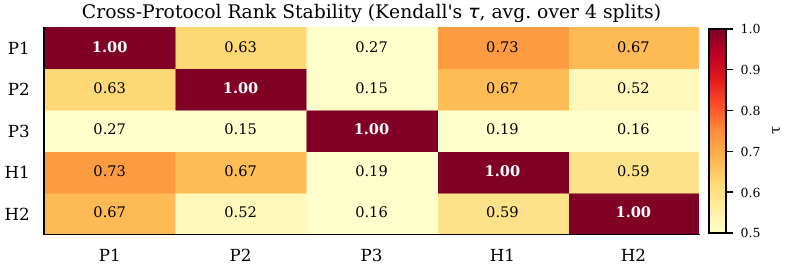}
  \caption{%
    \textbf{Cross-protocol rank stability (Kendall's $\tau$).} Pairwise agreement between model rankings induced by each evaluation protocol, averaged across cross-factor splits. Higher $\tau$ indicates that the two protocols yield consistent model orderings, while low values reveal protocol-sensitive ranking disagreements.
}
  \label{fig:cross_protocol_rank}
\end{figure}

\paragraph{Cross-protocol ranking stability}
As shown in 
Figure~\ref{fig:cross_protocol_rank}, model rankings are highly consistent across closed-set flat protocols (P1, P2, $\tau \ge 0.63$) indicating that these protocols largely probe a shared latent capability axis despite differences in prompting structure. Even structurally distant formulations such as direct forced-choice inference (P1) and event-grounded inference (P2) yield strong agreement, suggesting that protocol variations primarily modulate difficulty rather than relative model competence. In contrast, open-ended evaluation (P3) breaks this structure, exhibiting uniformly weak correlations with all closed-set protocols ($\tau \le 0.27$), consistent with a transition from constrained selection to unconstrained generation under LLM-based judging. This creates a clear separation between a stable structured-inference regime, where rankings are preserved across interventions, and an open-ended regime where evaluation dynamics induce reordering. Cross-protocol stability between flat and hierarchical protocols (P1$\leftrightarrow$H1: $\tau = 0.726$) further suggests these families probe a shared underlying capability axis; we return to this in Section \ref{subsec:hierar_insights}.



\begin{figure}[t]
  \centering
  \includegraphics[width=\columnwidth]{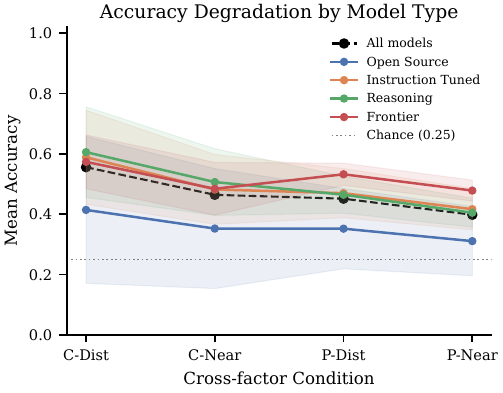}
  \caption{%
    \textbf{Accuracy degradation across the difficulty ladder.}
    Lines show mean accuracy per model category; shaded bands indicate
    $\pm 1$ standard deviation across models within each category.
  }
  \label{fig:degradation}
\end{figure}

\paragraph{Accuracy across difficulty regimes.}
We next consider the combined effect of typicality and distractor similarity across the full range of conditions in Figure~\ref{fig:degradation}
 under P1. Accuracy decreases as cross-factor proximity increases, but the degradation is structured rather than uniform, revealing a progressive compression of performance differences across model families.

At the easiest condition (C-Dist), clear separation is observed between model categories, with Frontier and Reasoning models reaching ~0.60 accuracy while Open Source models cluster near ~0.41. This separation steadily collapses with increasing difficulty, converging to a narrower band at P-Near, only marginally above chance. This indicates that joint difficulty erodes between-class separability, compressing performance toward a shared lower bound despite substantial initial capability gaps.

Beyond this global trend, Frontier models exhibit a non-monotonic pattern, with a partial recovery at P-Dist following an initial drop at C-Near, suggesting that scale may induce regime-specific advantages selectively triggered by particular combinations of typicality and distractor structure rather than yielding uniform robustness gains.

In contrast, Open Source models remain nearly flat and close to chance across conditions, consistent with an early saturation regime in which additional difficulty has limited effect on behavior. Finally, substantial within-category variance highlights that robustness under joint difficulty is primarily a model-level attribute rather than a category-level trait, with individual systems exhibiting markedly different degradation profiles.

\paragraph{Effect of event grounding.}
As shown in the top panels of Figure~\ref{fig:p1_p2_h1_h2}, introducing an explicit intermediate step yields sharply asymmetric effects across decomposition types. Event-grounded inference (P1$\rightarrow$P2) acts as a near-universal penalty, with consistent regressions across model families and difficulty levels. Notably, this degradation shows little dependence on baseline capability, suggesting that event grounding imposes a largely uniform “processing cost” rather than amplifying model strengths. See Appendix Figure \ref{fig:p1p2p3_heatmap} for full per-model accuracy across all splits and protocols.

\subsection{Hierarchical Abstraction}
\label{subsec:hierar_insights}
We examine whether hierarchical decomposition improves compositional inference, and how it shapes error dynamics and test-time reasoning.

\begin{figure}[t]
  \centering
  \includegraphics[width=\columnwidth]{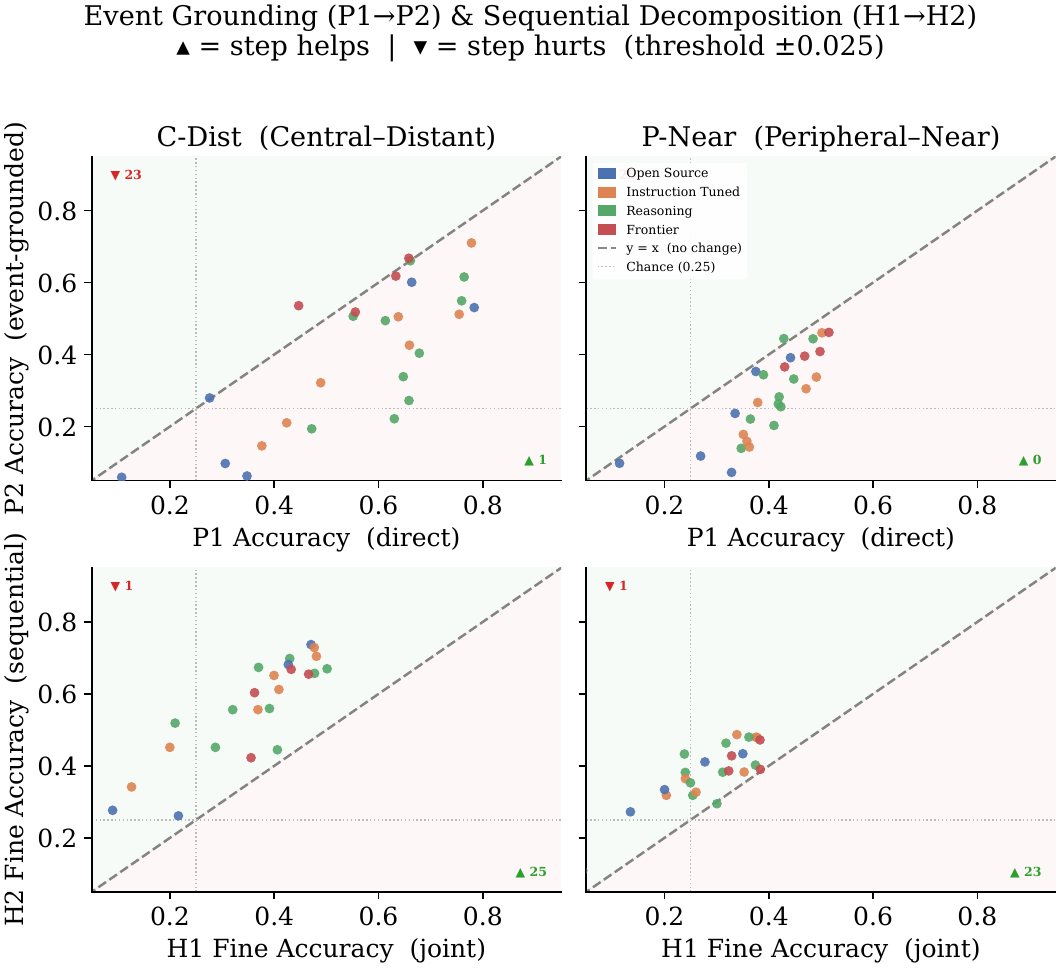}
  \caption{%
    \textbf{Event-grounded inference (P1→P2) and sequential hierarchical decomposition (H1→H2).}
    Top: per-model accuracy under direct activity inference (P1) vs.\ event-grounded inference (P2) for the easiest (Central-Distant) and hardest (Peripheral-Near) splits. Bottom: fine-grained accuracy under joint (H1) vs.\ sequential (H2) hierarchical abstraction for the same splits. Points above the diagonal indicate improvement of the two-stage approach over the single-stage baseline.
  }
  \label{fig:p1_p2_h1_h2}
\end{figure}

\paragraph{Sequential decomposition improves over joint}
In sharp contrast to event grounding, sequential hierarchical decomposition consistently improves fine-grained accuracy across nearly all models (Figure \ref{fig:p1_p2_h1_h2}, bottom panels), indicating that intermediate abstraction aligned with the task taxonomy can reduce effective search complexity without introducing additional failure modes. This holds across both easy (C-Dist) and hard (P-Near) conditions, with 25 out of 26 models improving in each split. The juxtaposition with the P1$\leftrightarrow$P2 regime reveals a structural distinction between two forms of decomposition: one that contracts the hypothesis space by conditioning fine-grained inference on a coarser semantic category (hierarchical labeling, H1$\leftrightarrow$H2), and one that adds an independent prediction requirement upstream of the main inference step (event grounding, P1$\leftrightarrow$P2). Only the former yields reliable gains. Intermediate deviations in the cross-protocol ranking matrix, such as H1$\leftrightarrow$H2 ($\tau = 0.594$), further indicate that sequential decomposition can perturb model orderings without collapsing the overall closed-set consistency established in Section \ref{subsec:results_scenario1}. Full per-model fine-grained accuracy across all splits and hierarchical protocols is reported in Appendix Figure \ref{fig:h1h2_heatmap}.

\begin{figure}[t]
  \centering
  \includegraphics[width=\columnwidth]{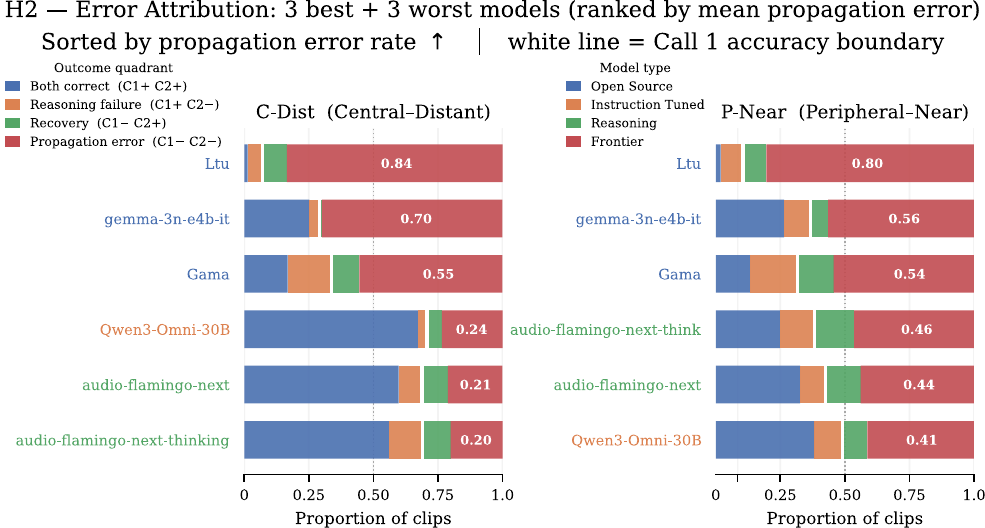}
    \caption{Error attribution for the three best and three worst models ranked by mean propagation error, shown for the C-Dist and P-Near conditions. Each bar decomposes model outputs into four outcome quadrants: both calls correct (C1$^+$C2$^+$), reasoning failure (C1$^+$C2$^-$), recovery (C1$^-$C2$^+$), and propagation error (C1$^-$C2$^-$). The white line sets the Call 1 accuracy boundary.}
  \label{fig:error_attribution}
\end{figure}

\paragraph{Error propagation and sequential failure modes.}
We next examine how errors propagate across sequential calls in Figure~\ref{fig:error_attribution}. A consistent pattern emerges: performance differences are primarily driven by failures in the first call, with markedly different capacities for downstream recovery.

Among the highest-error models (Ltu, gemma-3n-e4b-it, and Gama), performance is dominated by propagation error, indicating that initial misclassifications in Call 1 cascade almost entirely into Call 2 failures, with minimal evidence of recovery. Ltu represents an extreme case, exhibiting an almost binary failure structure with negligible recovery or correction. Gama shows a slightly more distributed error profile, with visible, but insufficient, reasoning failure and recovery components, indicating partial but ineffective error handling.

In contrast, the best-performing models achieve their advantage primarily through reduced Call 1 error rates rather than improved recovery mechanisms, indicating that successful sequential inference is largely determined upstream. Notably, the thinking variant of audio-flamingo-next does not consistently improve this dynamic: while differences are marginal in C-Dist, in P-Near it exhibits increased propagation error and a larger reasoning failure component, suggesting that extended reasoning may introduce drift rather than corrective structure in sequential audio settings.

Finally, moving from C-Dist to P-Near systematically increases propagation error across all models while compressing the performance gap between best and worst systems. This suggests a regime in which increased task difficulty amplifies early-stage failures and reduces the effective contribution of downstream inference, leading to a convergence toward similar failure modes across models.

\begin{figure}[t]
  \centering
  \includegraphics[width=\columnwidth]{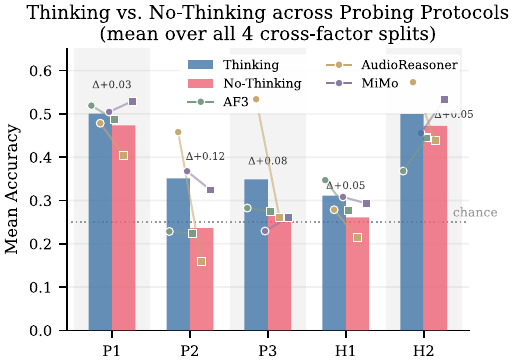}
  \caption{%
    Thinking vs.\ no-thinking accuracy across probing protocols and difficulty splits. Bars show mean accuracy, markers indicate paired model variants, and $\Delta$ denotes the mean protocol-wise difference.
  }
  \label{fig:thinking-vs-nothinking-by-protocol}
\end{figure}

\paragraph{Test-time reasoning (thinking vs.\ no-thinking).}
For model families offering both thinking and non-thinking variants (e.g., Audio Flamingo~3, Audio-Reasoner, MiMo-Audio; Figure \ref{fig:thinking-vs-nothinking-by-protocol}), extended test-time reasoning yields consistent but strongly protocol-dependent gains. Benefits are minimal in P1 ($\Delta\approx +0.03$), where tasks are largely solvable via direct pattern recognition, but increase substantially in P2 ($\Delta\approx +0.12$), where intermediate reasoning in the first stage compounds into improved inference under the two-step decomposition. H1 and H2 show intermediate gains ($\Delta\approx +0.05$), indicating partial advantage for hierarchical abstraction. Overall, thinking is most effective when the task supports intermediate abstraction and ambiguity resolution, but offers limited benefit when inference reduces to direct retrieval; notably, MiMo-Audio deviates from this trend, with its thinking variant providing only marginal improvement over the non-thinking version in H1.


\subsection{Human Performance}
\label{sec:human-validation}

To determine whether the model gap under Peripheral--Near conditions is attributable to a genuine capability limitation, we conducted a human evaluation on the easiest and hardest P1 conditions (Central--Distant and Peripheral--Near). Four annotators completed the same four-option forced-choice task as the models on a subset of 301 items (160 hard, 141 easy).

As shown in Table~\ref{tab:human-validation}, humans substantially
outperform models on both splits. On the easy split, the average human
annotator (0.864) exceeds the average model (0.320) by $+0.544$, and the
best model (0.780 - Qwen3-Omni) still trails majority-vote human performance (0.965) by
$+0.185$. On the hard split, the gap narrows in absolute terms but persists:
average human accuracy (0.580) exceeds average model accuracy (0.275) by
$+0.305$, and majority-vote humans (0.694) exceed the best model (0.500) by
$+0.194$. Humans outperform models on 96.6\% of easy items and 87.5\% of
hard items. Inter-annotator agreement is substantial (Fleiss' $\kappa =
0.797$ easy, $0.582$ hard).

Beyond aggregate accuracy, we compare which \emph{classes} humans and
models find difficult. Class-difficulty rankings correlate only moderately
between humans and models (Spearman's $\rho = 0.588$ easy, $0.565$ hard),
and overlap in the \emph{hardest}-ranked classes is especially low (1/5
easy, 1/5 hard), while overlap in the \emph{easiest}-ranked classes is
higher (5/5 easy, 3/5 hard). This indicates that models do not simply
underperform humans uniformly; they specifically misjudge acoustically
confusable activities that humans reliably distinguish, suggesting that the observed gap reflects a genuine limitation in compositional reasoning.

\begin{table}[t]
\centering
\small
\resizebox{\columnwidth}{!}{%
\begin{tabular}{lcccccc}
\toprule
\textbf{Split} & \textbf{Avg M} & \textbf{Best M} & \textbf{Avg H} &
\textbf{Maj H} & \textbf{$\Delta$Avg} & \textbf{H>M} \\
\midrule
Easy & 0.320 & 0.780 & 0.864 & 0.965 & +0.544 & 96.6\% \\
Hard & 0.275 & 0.500 & 0.580 & 0.694 & +0.305 & 87.5\% \\
\bottomrule
\end{tabular}%
}
\caption{Human vs.\ model accuracy on the easiest (Central--Distant) and
hardest (Peripheral--Near) P1 conditions. M = model, H = human, $\Delta$Avg
= human$-$model gain in average accuracy, H>M = \% of items where humans
outperform models.}
\label{tab:human-validation}
\end{table}

\section{Conclusion}
We introduced a benchmark and factorial evaluation framework probing a wide set of large audio-language models (LALMs) on compositional auditory understanding of human actions and activities. Our results expose three consistent findings: typicality robustness is largely decoupled from overall model capability; inferential decomposition matters critically, with event grounding uniformly hurting flat inference while taxonomy-aligned hierarchical decomposition reliably helps;
\added{and under joint difficulty, all model families converge near chance, revealing that current models do not reliably perform compositional inference from atomic acoustic events to higher-level human activities solely from audio. These findings suggest that such compositional reasoning does not emerge as a byproduct of general audio-language pretraining for everyday activity recognition from egocentric recordings.}
We release all data, outputs, and code to support progress on this open challenge on audio perception.

\newpage
\section*{Limitations and Future Work}
This work establishes a first controlled framework for compositional auditory understanding, and several directions naturally follow from its design choices. The single-label, fixed-duration segment setting provides a clean experimental foundation but opens the door to richer evaluations involving co-occurring and transitioning activities, and extended audio recordings. Exemplar typicality and distractor similarity are defined via CLAP embeddings, which offer a principled and reproducible difficulty axis. However, we did not cross-validate the CLAP-based typicality and similarity factors against an independent embedding model or a knowledge-grounded refinement of CLAP such as
iKnow-audio~\citep{olvera2025iknow}, which has been shown to reduce
CLAP's reliance on prompt engineering and improve disambiguation of
acoustically similar sounds. While our human evaluation
(\S\ref{sec:human-validation}) shows convergent behavioral evidence for
the resulting difficulty ordering, repeating exemplar and distractor
selection with such knowledge-refined embeddings would further strengthen
the construction and is a natural next step. 

The benchmark's egocentric sourcing reflects the availability of richly annotated activity corpora and suggests natural extension to broader cultural and linguistic contexts as such datasets mature. Finally, while LLM-based judging provides flexible semantic evaluation for open-ended protocols, future work could explore human judgement as an additional calibration signal. Together, these directions position the benchmark as a resource relevant to multimodal audio research.

\added{Importantly, our conclusions do not claim that our findings generalize to compositional auditory reasoning more broadly. Such reasoning may manifest differently in other domains, including music, speech, environmental soundscapes, or non-egocentric recordings, and evaluating these settings remains an important direction for future work.}

\section*{Acknowledgments}
This work was supported by Hi!Paris, State funding managed by the French National Research Agency (ANR) under the France 2030 program, reference ANR-23-IACL-0005, and by the European Union (ERC, HI-Audio, 101052978). Views and opinions expressed are however those of the author(s) only and do not necessarily reflect those of the European Union or the European Research
Council. Neither the European Union nor the granting authority can be held responsible for them.

\bibliography{custom}

\begin{thebibliography}{33}
\providecommand{\natexlab}[1]{#1}

\bibitem[{Chen et~al.(2020)Chen, Xie, Vedaldi, and Zisserman}]{Chen20}
Honglie Chen, Weidi Xie, Andrea Vedaldi, and Andrew Zisserman. 2020.
\newblock Vggsound: A large-scale audio-visual dataset.
\newblock In \emph{International Conference on Acoustics, Speech, and Signal Processing (ICASSP)}.

\bibitem[{Chu et~al.(2023)Chu, Xu, Zhou, Yang, Zhang, Yan, Zhou, and Zhou}]{chu2023qwen}
Yunfei Chu, Jin Xu, Xiaohuan Zhou, Qian Yang, Shiliang Zhang, Zhijie Yan, Chang Zhou, and Jingren Zhou. 2023.
\newblock Qwen-audio: Advancing universal audio understanding via unified large-scale audio-language models.
\newblock \emph{arXiv preprint arXiv:2311.07919}.

\bibitem[{Damen et~al.(2020)Damen, Doughty, Farinella, Fidler, Furnari, Kazakos, Moltisanti, Munro, Perrett, Price, and Wray}]{Damen2020Collection}
Dima Damen, Hazel Doughty, Giovanni~Maria Farinella, Sanja Fidler, Antonino Furnari, Evangelos Kazakos, Davide Moltisanti, Jonathan Munro, Toby Perrett, Will Price, and Michael Wray. 2020.
\newblock The epic-kitchens dataset: Collection, challenges and baselines.
\newblock \emph{IEEE Transactions on Pattern Analysis and Machine Intelligence (TPAMI)}.

\bibitem[{Demirel et~al.(2025)Demirel, Thakkar, Elizalde, Marques, Sarathy, Bai, Srinivas, Xu, Ren, and Narain}]{demirel2025using}
Ilker Demirel, Karan Thakkar, Benjamin Elizalde, Miquel~Espi Marques, Aditya Sarathy, Yang Bai, Umamahesh Srinivas, Jiajie Xu, Shirley Ren, and Jaya Narain. 2025.
\newblock Using llms for late multimodal sensor fusion for activity recognition.
\newblock \emph{arXiv preprint arXiv:2509.10729}.

\bibitem[{Deshmukh et~al.(2023)Deshmukh, Elizalde, Singh, and Wang}]{deshmukh2023pengi}
Soham Deshmukh, Benjamin Elizalde, Rita Singh, and Huaming Wang. 2023.
\newblock Pengi: An audio language model for audio tasks.
\newblock \emph{Advances in Neural Information Processing Systems}, 36:18090--18108.

\bibitem[{Elizalde et~al.(2023)Elizalde, Deshmukh, Al~Ismail, and Wang}]{elizalde2023clap}
Benjamin Elizalde, Soham Deshmukh, Mahmoud Al~Ismail, and Huaming Wang. 2023.
\newblock Clap learning audio concepts from natural language supervision.
\newblock In \emph{International Conference on Acoustics, Speech, and Signal Processing (ICASSP)}, pages 1--5. IEEE.

\bibitem[{Fonseca et~al.(2022)Fonseca, Favory, Pons, Font, and Serra}]{fonseca2022FSD50K}
Eduardo Fonseca, Xavier Favory, Jordi Pons, Frederic Font, and Xavier Serra. 2022.
\newblock {FSD50K}: an open dataset of human-labeled sound events.
\newblock \emph{IEEE/ACM Transactions on Audio, Speech, and Language Processing}, 30:829--852.

\bibitem[{Gemmeke et~al.(2017)Gemmeke, Ellis, Freedman, Jansen, Lawrence, Moore, Plakal, and Ritter}]{elisAudioset}
Jort~F. Gemmeke, Daniel P.~W. Ellis, Dylan Freedman, Aren Jansen, Wade Lawrence, R.~Channing Moore, Manoj Plakal, and Marvin Ritter. 2017.
\newblock Audio set: An ontology and human-labeled dataset for audio events.
\newblock In \emph{International Conference on Acoustics, Speech, and Signal Processing (ICASSP)}.

\bibitem[{Ghosh et~al.(2026)Ghosh, Goel, Kim, Kumar, Kong, gil Lee, Yang, Duraiswami, Manocha, Valle, and Catanzaro}]{goel2025audio}
Sreyan Ghosh, Arushi Goel, Jaehyeon Kim, Sonal Kumar, Zhifeng Kong, Sang gil Lee, Chao-Han~Huck Yang, Ramani Duraiswami, Dinesh Manocha, Rafael Valle, and Bryan Catanzaro. 2026.
\newblock Audio flamingo 3: Advancing audio intelligence with fully open large audio language models.
\newblock In \emph{The Thirty-ninth Annual Conference on Neural Information Processing Systems}.

\bibitem[{Ghosh et~al.(2025)Ghosh, Kong, Kumar, Sakshi, Kim, Ping, Valle, Manocha, and Catanzaro}]{ghosh2025audio}
Sreyan Ghosh, Zhifeng Kong, Sonal Kumar, S~Sakshi, Jaehyeon Kim, Wei Ping, Rafael Valle, Dinesh Manocha, and Bryan Catanzaro. 2025.
\newblock Audio flamingo 2: An audio-language model with long-audio understanding and expert reasoning abilities.
\newblock In \emph{Forty-second International Conference on Machine Learning}.

\bibitem[{Ghosh et~al.(2024)Ghosh, Seth, Kumar, Tyagi, Evuru, S, Sakshi, Nieto, Duraiswami, and Manocha}]{ghosh2023compa}
Sreyan Ghosh, Ashish Seth, Sonal Kumar, Utkarsh Tyagi, Chandra Kiran~Reddy Evuru, Ramaneswaran S, S~Sakshi, Oriol Nieto, Ramani Duraiswami, and Dinesh Manocha. 2024.
\newblock Compa: Addressing the gap in compositional reasoning in audio-language models.
\newblock In \emph{The Twelfth International Conference on Learning Representations}.

\bibitem[{Gong et~al.(2024)Gong, Luo, Liu, Karlinsky, and Glass}]{gong2023listen}
Yuan Gong, Hongyin Luo, Alexander~H. Liu, Leonid Karlinsky, and James~R. Glass. 2024.
\newblock Listen, think, and understand.
\newblock In \emph{The Twelfth International Conference on Learning Representations}.

\bibitem[{Grattafiori et~al.(2024)Grattafiori, Dubey, Jauhri, Pandey, Kadian, Al-Dahle, Letman, Mathur, Schelten, Vaughan et~al.}]{grattafiori2024llama}
Aaron Grattafiori, Abhimanyu Dubey, Abhinav Jauhri, Abhinav Pandey, Abhishek Kadian, Ahmad Al-Dahle, Aiesha Letman, Akhil Mathur, Alan Schelten, Alex Vaughan, and 1 others. 2024.
\newblock The llama 3 herd of models.
\newblock \emph{arXiv preprint arXiv:2407.21783}.

\bibitem[{Grauman et~al.(2022)Grauman, Westbury, Byrne, Chavis, Furnari, Girdhar, Hamburger, Jiang, Liu, Liu, Martin, Nagarajan, Radosavovic, Ramakrishnan, Ryan, Sharma, Wray, Xu, Xu, Zhao, Bansal, Batra, Cartillier, Crane, Do, Doulaty, Erapalli, Feichtenhofer, Fragomeni, Fu, Gebreselasie, Gonz\'alez, Hillis, Huang, Huang, Jia, Khoo, Kol\'a\v{r}, Kottur, Kumar, Landini, Li, Li, Li, Mangalam, Modhugu, Munro, Murrell, Nishiyasu, Price, Ruiz, Ramazanova, Sari, Somasundaram, Southerland, Sugano, Tao, Vo, Wang, Wu, Yagi, Zhao, Zhu, Arbel\'aez, Crandall, Damen, Farinella, Fuegen, Ghanem, Ithapu, Jawahar, Joo, Kitani, Li, Newcombe, Oliva, Park, Rehg, Sato, Shi, Shou, Torralba, Torresani, Yan, and Malik}]{Grauman_2022_CVPR}
Kristen Grauman, Andrew Westbury, Eugene Byrne, Zachary Chavis, Antonino Furnari, Rohit Girdhar, Jackson Hamburger, Hao Jiang, Miao Liu, Xingyu Liu, Miguel Martin, Tushar Nagarajan, Ilija Radosavovic, Santhosh~Kumar Ramakrishnan, Fiona Ryan, Jayant Sharma, Michael Wray, Mengmeng Xu, Eric~Zhongcong Xu, and 66 others. 2022.
\newblock Ego4d: Around the world in 3,000 hours of egocentric video.
\newblock In \emph{Proceedings of the IEEE/CVF Conference on Computer Vision and Pattern Recognition (CVPR)}, pages 18995--19012.

\bibitem[{Grauman et~al.(2024)Grauman, Westbury, Torresani, Kitani, Malik, Afouras, Ashutosh, Baiyya, Bansal, Boote, Byrne, Chavis, Chen, Cheng, Chu, Crane, Dasgupta, Dong, Escobar, Forigua, Gebreselasie, Haresh, Huang, Islam, Jain, Khirodkar, Kukreja, Liang, Liu, Majumder, Mao, Martin, Mavroudi, Nagarajan, Ragusa, Ramakrishnan, Seminara, Somayazulu, Song, Su, Xue, Zhang, Zhang, Castillo, Chen, Fu, Furuta, Gonzalez, Gupta, Hu, Huang, Huang, Khoo, Kumar, Kuo, Lakhavani, Liu, Luo, Luo, Meredith, Miller, Oguntola, Pan, Peng, Pramanick, Ramazanova, Ryan, Shan, Somasundaram, Song, Southerland, Tateno, Wang, Wang, Yagi, Yan, Yang, Yu, Zha, Zhao, Zhao, Zhu, Zhuo, Arbelaez, Bertasius, Damen, Engel, Farinella, Furnari, Ghanem, Hoffman, Jawahar, Newcombe, Park, Rehg, Sato, Savva, Shi, Shou, and Wray}]{Grauman_2024_CVPRexo}
Kristen Grauman, Andrew Westbury, Lorenzo Torresani, Kris Kitani, Jitendra Malik, Triantafyllos Afouras, Kumar Ashutosh, Vijay Baiyya, Siddhant Bansal, Bikram Boote, Eugene Byrne, Zach Chavis, Joya Chen, Feng Cheng, Fu-Jen Chu, Sean Crane, Avijit Dasgupta, Jing Dong, Maria Escobar, and 81 others. 2024.
\newblock Ego-exo4d: Understanding skilled human activity from first- and third-person perspectives.
\newblock In \emph{Proceedings of the IEEE/CVF Conference on Computer Vision and Pattern Recognition (CVPR)}, pages 19383--19400.

\bibitem[{Guzhov et~al.(2022)Guzhov, Raue, Hees, and Dengel}]{guzhov2022audioclip}
Andrey Guzhov, Federico Raue, J{\"o}rn Hees, and Andreas Dengel. 2022.
\newblock Audioclip: Extending clip to image, text and audio.
\newblock In \emph{International Conference on Acoustics, Speech, and Signal Processing (ICASSP)}, pages 976--980. IEEE.

\bibitem[{Kong et~al.(2024)Kong, Goel, Badlani, Ping, Valle, and Catanzaro}]{kong2024audio}
Zhifeng Kong, Arushi Goel, Rohan Badlani, Wei Ping, Rafael Valle, and Bryan Catanzaro. 2024.
\newblock Audio flamingo: A novel audio language model with few-shot learning and dialogue abilities.
\newblock In \emph{International Conference on Machine Learning}, pages 25125--25148. PMLR.

\bibitem[{Kumar et~al.(2026)Kumar, Sedl{\'a}{\v{c}}ek, Lokegaonkar, L{\'o}pez, Yu, Anand, Ryu, Chen, Pli{\v{c}}ka, Hlav{\'a}{\v{c}}ek et~al.}]{kumar2026mmau}
Sonal Kumar, {\v{S}}imon Sedl{\'a}{\v{c}}ek, Vaibhavi Lokegaonkar, Fernando L{\'o}pez, Wenyi Yu, Nishit Anand, Hyeonggon Ryu, Lichang Chen, Maxim Pli{\v{c}}ka, Miroslav Hlav{\'a}{\v{c}}ek, and 1 others. 2026.
\newblock Mmau-pro: A challenging and comprehensive benchmark for holistic evaluation of audio general intelligence.
\newblock In \emph{Proceedings of the AAAI Conference on Artificial Intelligence}.

\bibitem[{Li et~al.(2025{\natexlab{a}})Li, Liu, Dinkel, Niu, Zhang, and Luan}]{li2025reinforcement}
Gang Li, Jizhong Liu, Heinrich Dinkel, Yadong Niu, Junbo Zhang, and Jian Luan. 2025{\natexlab{a}}.
\newblock Reinforcement learning outperforms supervised fine-tuning: A case study on audio question answering.
\newblock \emph{arXiv preprint arXiv:2503.11197}.

\bibitem[{Li et~al.(2025{\natexlab{b}})Li, Liu, Zhang, Chen, Li, Li, Liu, Ming, Dong, Pan et~al.}]{li2025baichuan}
Yadong Li, Jun Liu, Tao Zhang, Song Chen, Tianpeng Li, Zehuan Li, Lijun Liu, Lingfeng Ming, Guosheng Dong, Da~Pan, and 1 others. 2025{\natexlab{b}}.
\newblock Baichuan-omni-1.5 technical report.
\newblock \emph{arXiv preprint arXiv:2501.15368}.

\bibitem[{Ma et~al.(2025{\natexlab{a}})Ma, Chen, Wang, Chng, and Chen}]{ma2025audio}
Ziyang Ma, Zhuo Chen, Yuping Wang, Eng~Siong Chng, and Xie Chen. 2025{\natexlab{a}}.
\newblock Audio-cot: Exploring chain-of-thought reasoning in large audio language model.
\newblock \emph{arXiv preprint arXiv:2501.07246}.

\bibitem[{Ma et~al.(2025{\natexlab{b}})Ma, Ma, Zhu, Yang, Chao, Xu, Chen, Chen, Chen, Cong et~al.}]{ma2025mmar}
Ziyang Ma, Yinghao Ma, Yanqiao Zhu, Chen Yang, Yi-Wen Chao, Ruiyang Xu, Wenxi Chen, Yuanzhe Chen, Zhuo Chen, Jian Cong, and 1 others. 2025{\natexlab{b}}.
\newblock Mmar: A challenging benchmark for deep reasoning in speech, audio, music, and their mix.
\newblock \emph{arXiv preprint arXiv:2505.13032}.

\bibitem[{Olvera et~al.(2024)Olvera, Stamatiadis, and Essid}]{Olvera2024description}
Michel Olvera, Paraskevas Stamatiadis, and Slim Essid. 2024.
\newblock A sound description: Exploring prompt templates and class descriptions to enhance zero-shot audio classification.
\newblock In \emph{Proceedings of the Detection and Classification of Acoustic Scenes and Events (DCASE)}, pages 116--120.

\bibitem[{Olvera et~al.(2025)Olvera, Wang, Stamatiadis, Richard, and Essid}]{olvera2025iknow}
Michel Olvera, Changhong Wang, Paraskevas Stamatiadis, Ga{\"e}l Richard, and Slim Essid. 2025.
\newblock iknow-audio: Integrating knowledge graphs with audio-language models.
\newblock In \emph{Proceedings of the 2025 Conference on Empirical Methods in Natural Language Processing}, pages 34671--34688.

\bibitem[{Perrett et~al.(2025)Perrett, Darkhalil, Sinha, Emara, Pollard, Parida, Liu, Gatti, Bansal, Flanagan, Chalk, Zhu, Guerrier, Abdelazim, Zhu, Moltisanti, Wray, Doughty, and Damen}]{perrett2025hdepic}
Toby Perrett, Ahmad Darkhalil, Saptarshi Sinha, Omar Emara, Sam Pollard, Kranti Parida, Kaiting Liu, Prajwal Gatti, Siddhant Bansal, Kevin Flanagan, Jacob Chalk, Zhifan Zhu, Rhodri Guerrier, Fahd Abdelazim, Bin Zhu, Davide Moltisanti, Michael Wray, Hazel Doughty, and Dima Damen. 2025.
\newblock Hd-epic: A highly-detailed egocentric video dataset.
\newblock In \emph{Proceedings of the IEEE/CVF Conference on Computer Vision and Pattern Recognition (CVPR)}.

\bibitem[{Piczak(2015)}]{piczak2015dataset}
Karol~J Piczak. 2015.
\newblock Esc: Dataset for environmental sound classification.
\newblock In \emph{Proceedings of the 23rd ACM international conference on Multimedia}, pages 1015--1018.

\bibitem[{Sakshi et~al.(2025)Sakshi, Tyagi, Kumar, Seth, Selvakumar, Nieto, Duraiswami, Ghosh, and Manocha}]{sakshi2025mmau}
Sakshi Sakshi, Utkarsh Tyagi, Sonal Kumar, Ashish Seth, Ramaneswaran Selvakumar, Oriol Nieto, Ramani Duraiswami, Sreyan Ghosh, and Dinesh Manocha. 2025.
\newblock Mmau: A massive multi-task audio understanding and reasoning benchmark.
\newblock In \emph{International Conference on Learning Representations}.

\bibitem[{Stamatiadis et~al.(2024)Stamatiadis, Olvera, and Essid}]{Stamatiadis2024}
Paraskevas Stamatiadis, Michel Olvera, and Slim Essid. 2024.
\newblock Salt: Standardized audio event label taxonomy.
\newblock In \emph{Proceedings of the Detection and Classification of Acoustic Scenes and Events (DCASE)}.

\bibitem[{Tang et~al.(2024)Tang, Yu, Sun, Chen, Tan, Li, Lu, MA, and Zhang}]{tang2023salmonn}
Changli Tang, Wenyi Yu, Guangzhi Sun, Xianzhao Chen, Tian Tan, Wei Li, Lu~Lu, Zejun MA, and Chao Zhang. 2024.
\newblock {SALMONN}: Towards generic hearing abilities for large language models.
\newblock In \emph{The Twelfth International Conference on Learning Representations}.

\bibitem[{Wijngaard et~al.(2026)Wijngaard, Formisano, Esposito, and Dumontier}]{wijngaard2025audsemthinker}
Gijs Wijngaard, Elia Formisano, Michele Esposito, and Michel Dumontier. 2026.
\newblock Audsemthinker: Enhancing audio-language models through reasoning over semantics of sound.
\newblock In \emph{The Thirty-ninth Annual Conference on Neural Information Processing Systems}.

\bibitem[{Wu et~al.(2023)Wu, Chen, Zhang, Hui, Berg-Kirkpatrick, and Dubnov}]{wu2023large}
Yusong Wu, Ke~Chen, Tianyu Zhang, Yuchen Hui, Taylor Berg-Kirkpatrick, and Shlomo Dubnov. 2023.
\newblock Large-scale contrastive language-audio pretraining with feature fusion and keyword-to-caption augmentation.
\newblock In \emph{International Conference on Acoustics, Speech, and Signal Processing (ICASSP)}, pages 1--5. IEEE.

\bibitem[{Xu et~al.(2025)Xu, Guo, He, Hu, He, Bai, Chen, Wang, Fan, Dang, Zhang, Wang, Chu, and Lin}]{Qwen2.5-Omni}
Jin Xu, Zhifang Guo, Jinzheng He, Hangrui Hu, Ting He, Shuai Bai, Keqin Chen, Jialin Wang, Yang Fan, Kai Dang, Bin Zhang, Xiong Wang, Yunfei Chu, and Junyang Lin. 2025.
\newblock \href {https://arxiv.org/abs/2503.20215} {Qwen2.5-omni technical report}.
\newblock \emph{Preprint}, arXiv:2503.20215.

\bibitem[{Zhifei et~al.(2025)Zhifei, Lin, Liu, Wu, Yan, and Miao}]{xie2025audio}
Xie Zhifei, Mingbao Lin, Zihang Liu, Pengcheng Wu, Shuicheng Yan, and Chunyan Miao. 2025.
\newblock Audio-reasoner: Improving reasoning capability in large audio language models.
\newblock In \emph{Proceedings of the 2025 Conference on Empirical Methods in Natural Language Processing}, pages 23840--23862.

\end{thebibliography}
\newpage\newpage

\appendix
\section{Appendix}
\label{sec:appendix}

\subsection{Two-Round Audibility Filtering}
\label{sec:audibility_filtering}

\paragraph{First round: action audibility screening.}
Each candidate action label is assigned to one of four categories based on whether the action can, in principle, be recognized from sound alone. Only categories 3 and 4 advance to the second pass. This stage is motivated by the need to construct an audio-grounded taxonomy: many egocentric action labels describe visually defined events that are not consistently reflected in the audio stream. We therefore filter labels to retain only those with stable and discriminative acoustic evidence. Table \ref{tab:first_pass_audibility} describes the screening categories.

\begin{table}[h]
\centering
\small
\begin{tabularx}{\columnwidth}{c X c}
\hline
\textbf{Category} & \textbf{Meaning} & \textbf{Decision} \\
\hline
1 & Predominantly visual; not useful for audio (e.g.\ \textit{moving\_baby}, \textit{reading}, \textit{writing}) & Reject \\
2 & Too specific or too brief to yield consistent audio signal (e.g.\ \textit{archery}) & Reject \\
3 & Ambiguous, but plausibly audible (e.g.\ \textit{acting\_in\_play}) & Retain \\
4 & Clearly and reliably audible (e.g.\ \textit{applauding}, \textit{drilling}, \textit{playing piano}) & Retain \\
\hline
\end{tabularx}
\caption{First-pass audibility screening categories.}
\label{tab:first_pass_audibility}
\end{table}

\paragraph{Second round: audibility verification.}
For each surviving label, between one and two minutes of audio are sampled across six to fourteen distinct source files. The label is then rated according to how reliably the action can be identified from audio alone, with the goal of retaining only labels with consistent acoustic evidence. Table \ref{tab:second_pass_audibility} describes the verification categories.

\begin{table}[h]
\centering
\small
\begin{tabularx}{\columnwidth}{c X c}
\hline
\textbf{Category} & \textbf{Meaning} & \textbf{Decision} \\
\hline
1 & Very poor --- sound absent or uninformative & Reject \\
2 & Poor --- audible events present but insufficient to detect the action from sound alone & Retain \\
3 & Good --- action clearly recognisable from sound & Retain \\
4 & Very good --- action acoustically dominant and unambiguous & Retain \\
\hline
\end{tabularx}
\caption{Second-pass audibility verification categories.}
\label{tab:second_pass_audibility}
\end{table}

Only category 1 is discarded, as it indicates that the target action is not reliably recoverable from the audio. Categories 2--4 are retained to preserve actions with varying but still informative acoustic evidence. A label must survive both passes to enter the taxonomy.

\subsection{Prompt Bank}
\label{app:prompt_bank}

Surface-level prompt variation can substantially influence model performance~\cite{Olvera2024description}. To control for sensitivity to linguistic surface forms, we use a diversified prompt bank spanning three stylistic families: (A) direct classification, (B) instructional, and (C) cloze/multiple-choice. Closed-set protocols use 15 disjoint templates, hierarchical abstraction protocols use 5 templates, and open-ended evaluation uses 10 templates. Examples from each family are shown below:

\paragraph{Direct classification}

\begin{quote}\small
\textit{What action is taking place in this audio? Pick one label from \texttt{\{LABELS\}}. Respond with only the label.}
\end{quote}

\begin{quote}\small
\textit{From the list \texttt{\{LABELS\}}, which activity best matches this recording. Output the label only.}
\end{quote}

\begin{quote}\small
\textit{Identify the activity in the clip. Your options are: \texttt{\{LABELS\}}. Respond with a single label.}
\end{quote}

\paragraph{Instructional}

\begin{quote}\small
\textit{Choose one label from the following: \texttt{\{LABELS\}}. Do not explain. Return only a label from the list.}
\end{quote}

\begin{quote}\small
\textit{This audio corresponds to which activity? Pick one option from \texttt{\{LABELS\}}. Give only a label. Return only a label from the list.}
\end{quote}

\begin{quote}\small
\textit{Which single activity best fits this sound? Pick from \texttt{\{LABELS\}}. Output only the label. Return only a label from the list.}
\end{quote}

\paragraph{Cloze/Multiple-choice}

\begin{quote}\small
\textit{The activity in this audio is: \_\_\_. Options: \texttt{\{LABELS\}}. Fill the blank with exactly one option.}
\end{quote}

\begin{quote}\small
\textit{Among the possible activities \texttt{\{LABELS\}}, which is the best match ? Output exactly one of them.}
\end{quote}

\begin{quote}\small
\textit{Audio classification task: Pick the best-fitting label from \texttt{\{LABELS\}}. Give the label as your complete answer.}
\end{quote}
The full set of prompts can be found in the data resources in the companion website.

\subsection{Full list of evaluated models}
\label{app:evaluated_models}
\paragraph{Open-source generative ALMs}: \textit{
LTU, GAMA, SALMONN, Qwen3-Omni-30B-A3B-Captioner, Audio Flamingo~Next-Captioner, Gemma-3n-E4B-it}.
\paragraph{Instruction-tuned ALMs}: \textit{Kimi-Audio-7B-Instruct, Qwen2-Audio-7B-Instruct, Qwen2.5-Omni-3B/7B, Qwen3-Omni-30B-A3B-Instruct, DeSTA2.5-Audio, Phi-4-Multimodal-Instruct}.
\paragraph{Reasoning-augmented ALMs}: \textit{Audio Flamingo~3 (Thinking / No Thinking), Audio Flamingo~Next (Thinking / No Thinking), Audio-Reasoner (Thinking / No Thinking), MiMo-Audio (Thinking / No Thinking), AudSemThinker, Qwen3-Omni-30B-A3B-Thinking, Gemma-4-E4B-it}.
\paragraph{Proprietary frontier models}: \textit{Gemini~2.0-Flash-Lite, Gemini~2.5-Flash-Lite, Gemini~2.0-Flash, Gemini~2.5-Flash}.

\subsection{LLM-as-Judge Evaluation}
\label{app:llm_judge}

Judgements are produced by \texttt{Meta-Llama-3.1-8B-Instruct}~\citep{grattafiori2024llama}, loaded in \texttt{bfloat16} precision via the Hugging Face \texttt{transformers} \texttt{pipeline} API with \texttt{device\_map="auto"}. Generation is near-greedy (\texttt{temperature}~$=0.1$, \texttt{do\_sample=True}, \texttt{max\_new\_tokens}~$=10$).

Two judging prompts were used, one lexical prompt for the closed-set protocols (P1, P2, H1, H2) and a separate semantic prompt for the open-ended protocol (P3).

\paragraph{Closed-set judging prompt (P1, P2, H1, H2))}
\begin{quote}\small
\textit{You are an evaluator for an audio-language model.
The ground-truth label for the activity is: ``\textlangle ground\_truth\textrangle''.
The model answered: ``\textlangle model\_output\textrangle''. Decide if the model's answer refers to the same activity as the ground-truth.\\
-- Respond ``Correct'' ONLY if the model output is lexically close to the
ground-truth (allowing for small spelling mistakes or typos).\\
-- Respond ``Incorrect'' if the model output refers to a different activity,
is too vague, or unrelated.\\
Output only ``Correct'' or ``Incorrect''.}
\end{quote}
\paragraph{Open-set judging prompt (P3)}
\begin{quote}\small
\textit{You are an evaluator for an audio-language model. The ground-truth label for the activity is:  ``\textlangle ground\_truth\textrangle''. The model answered: ``\textlangle model\_output\textrangle''.\\
--  Respond ``Correct'' if the model output is a valid synonym, paraphrase, or closely matching description of the ground-truth activity.\\
-- Respond ``Incorrect''  if the model output refers to a different activity, is too vague, or unrelated.\\
Output only ``Correct'' or ``Incorrect''.}
\end{quote}

\paragraph{Scoring.}
The LLM-as-Judge score for a given model, benchmark split, and prompt template
is the proportion of items receiving a \textit{Correct} judgement:
$$
\mathrm{LJ} = \frac{|\{\,i : \text{judge}(i) = \textit{Correct}\,\}|}{N}
$$
where $N$ is the total number of evaluated clips.
The judge is applied to the model's raw free-form output (\texttt{\{model\}\_activity}) against the canonical taxonomy label.

\section{Data and Reproducibility}
The source audio originates from Ego4D and Ego-Exo4D and cannot be redistributed due to their licensing terms. To support reproducibility within these constraints, we release all downstream benchmark artifacts, including benchmark metadata (take IDs, timestamps, labels), the taxonomy, prompts, distractors, model predictions, LLM-judge outputs, and human validation results, enabling all reported figures and tables to be recomputed without access to the original media. For researchers with licensed access to the source datasets (granted through free registration), we additionally release deterministic benchmark construction code that reconstructs the benchmark from the released take ID/timestamp pairs and reproduces the full evaluation pipeline end-to-end.

\section{Dataset Licenses}
The source datasets used to construct the benchmark taxonomy and audio exemplars are subject to the following licenses and terms of use. Ego4D is released under the Ego4D License Agreement, which permits non-commercial research use. Ego-Exo4D is distributed under the same terms. EPIC-KITCHENS is available under a Creative Commons Attribution 4.0 International (CC BY 4.0) license. HD-EPIC follows the terms of use of its parent corpus, EPIC-KITCHENS. All datasets are used strictly within their intended research-use scope: no clips are redistributed, and all experiments are conducted for non-commercial academic evaluation purposes only. The benchmark assets hosted on the companion website are derived from these sources and are made available under equivalent non-commercial research terms.

\subsection{Use of AI Assistants}
The authors used large language model-based assistants during the preparation of this manuscript, strictly for writing assistance tasks such as grammar correction, sentence rephrasing, and prose clarity improvements. No AI assistant was used to generate experimental results, produce or modify code for the evaluation framework, or make scientific claims. All scientific content, experimental design, analysis, and conclusions are entirely the authors' own.

\newpage

\begin{figure*}[t]
    \centering
    \includegraphics[width=\linewidth]{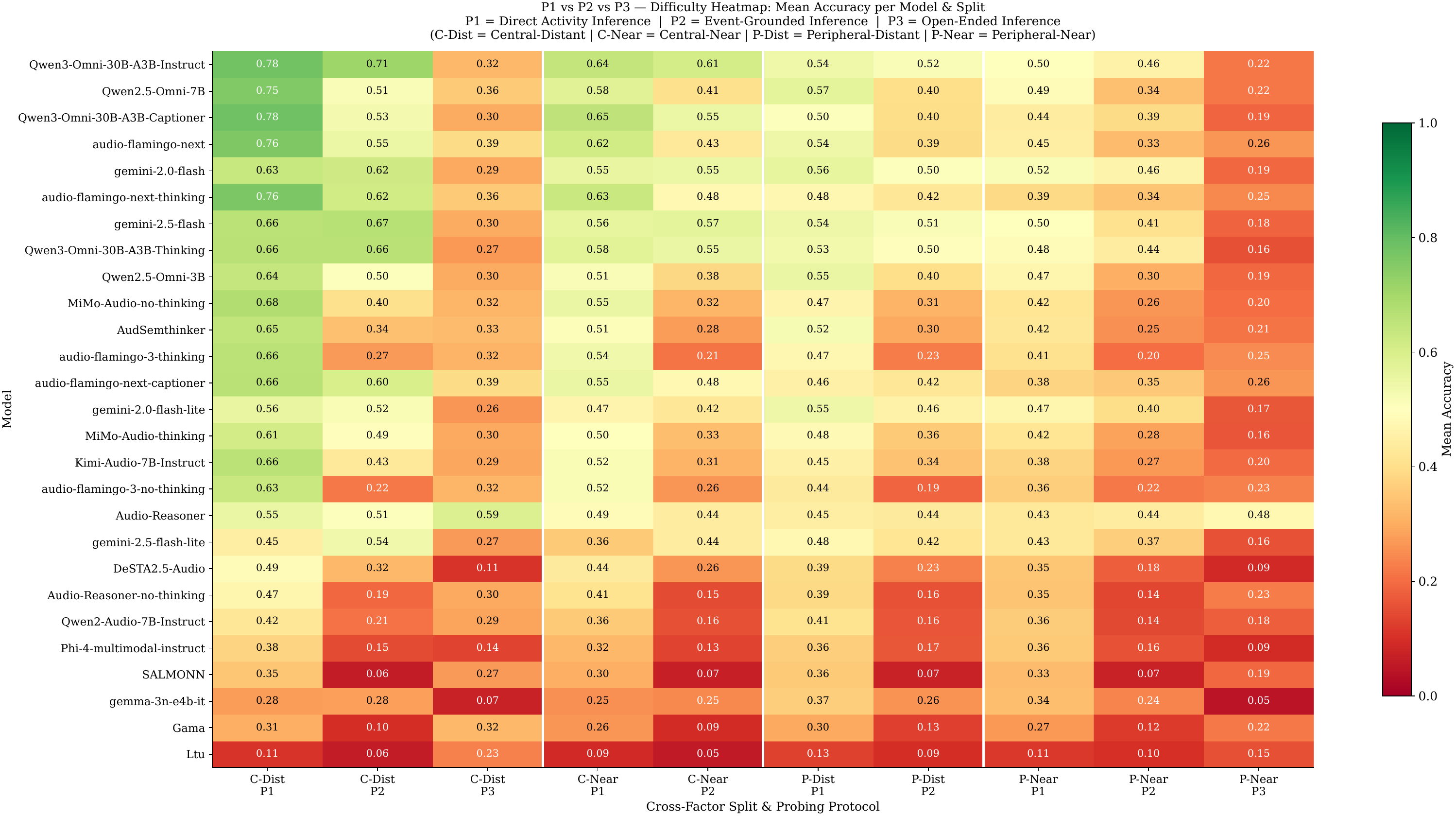}
    \caption{Mean accuracy per model across the four difficulty splits (C-Dist, C-Near, P-Dist, P-Near) for all three Probe Family~I protocols: \textbf{P1} (direct forced-choice), \textbf{P2} (event-grounded two-stage), and \textbf{P3} (open-ended LLM-judged). Columns are interleaved within each split; models are sorted by mean P1 accuracy (descending).}
    \label{fig:p1p2p3_heatmap}
\end{figure*}

\begin{figure*}[h!]
    \centering
    \includegraphics[width=\linewidth]{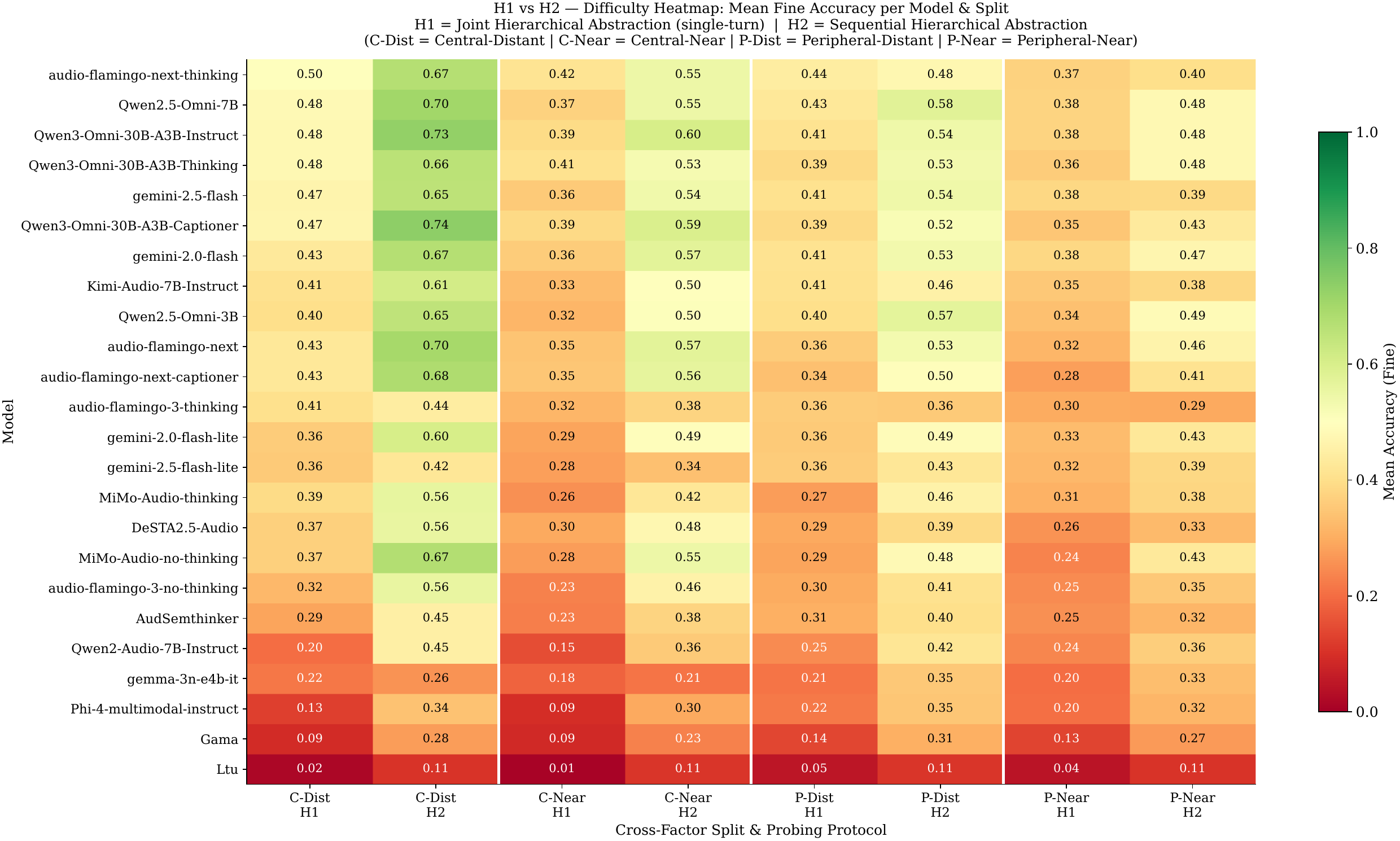}
    \caption{Mean fine-grained accuracy per model across the four difficulty splits (C-Dist, C-Near, P-Dist, P-Near) for Probe Family~II protocols: \textbf{H1} (joint hierarchical abstraction, single-turn) and \textbf{H2} (sequential hierarchical abstraction, multi-turn with audio cue). Columns are interleaved within each split; models are sorted by mean H1 accuracy (descending).}
    \label{fig:h1h2_heatmap}
\end{figure*}

\end{document}